\documentclass[12pt]{article}
\usepackage{graphicx,psfrag,epsf}
\usepackage{enumerate}
\usepackage{natbib}

\usepackage[margin=1in]{geometry}
\newcommand{\blind}{0}

\usepackage{amsmath,amsthm,amssymb}
\usepackage{mathrsfs}
\usepackage{filemod}
\usepackage[breaklinks=true,hidelinks]{hyperref}
\usepackage{array}

\usepackage{subfig}

\usepackage{indentfirst}

\def\note#1{\par\smallskip%
\noindent\kern-0.01\hsize%
\setlength\fboxrule{0pt}\fbox{\setlength\fboxrule{0.5pt}\fbox{%
\llap{$\boldsymbol\Longrightarrow$ }%
\vtop{\hsize=0.98\hsize\parindent=0cm\small\rm #1}%
\rlap{$\enskip\,\boldsymbol\Longleftarrow$}
}}%
}

\usepackage{color} 
\usepackage{booktabs} 

\def\ben#1{\begin{equation}#1\end{equation}}

\renewcommand{\phi}{\varphi}
\newcommand{\eps}{\varepsilon}

\def\given{\typeout{Command 'given' should only be used within bracket command}}
\newcounter{@bracketlevel}
\def\@bracketfactory#1#2#3#4#5#6{
	\expandafter\def\csname#1\endcsname##1{%
		\addtocounter{@bracketlevel}{1}%
		\global\expandafter\let\csname @middummy\alph{@bracketlevel}\endcsname\given%
		\global\def\given{\mskip#5\csname#4\endcsname\vert\mskip#6}\csname#4l\endcsname#2##1\csname#4r\endcsname#3%
		\global\expandafter\let\expandafter\given\csname @middummy\alph{@bracketlevel}\endcsname
		\addtocounter{@bracketlevel}{-1}}%
}
\def\bracketfactory#1#2#3{%
	\@bracketfactory{#1}{#2}{#3}{relax}{1mu plus 0.25mu minus 0.25mu}{0.6mu plus 0.15mu minus 0.15mu}
	\@bracketfactory{b#1}{#2}{#3}{big}{1mu plus 0.25mu minus 0.25mu}{0.6mu plus 0.15mu minus 0.15mu}
	\@bracketfactory{bb#1}{#2}{#3}{Big}{2.4mu plus 0.8mu minus 0.8mu}{1.8mu plus 0.6mu minus 0.6mu}
	\@bracketfactory{bbb#1}{#2}{#3}{bigg}{3.2mu plus 1mu minus 1mu}{2.4mu plus 0.75mu minus 0.75mu}
	\@bracketfactory{bbbb#1}{#2}{#3}{Bigg}{4mu plus 1mu minus 1mu}{3mu plus 0.75mu minus 0.75mu}
}
\bracketfactory{ceil}{\lceil}{\rceil}
\bracketfactory{abs}{\lvert}{\rvert}
\bracketfactory{clr}{(}{)}
\def\tbf#1{\textbf{#1}}

\usepackage[acronym,automake]{glossaries}
\newacronym{acr:mae}{MAE}{mean absolute error}
\newacronym{acr:mse}{MSE}{mean squared error}
\newacronym{acr:rmse}{RMSE}{root mean squared error}
\newacronym{acr:sw}{SW}{step-wise}
\makeglossaries

\def\mline#1{\begin{tabular}{@{}r@{}}#1\end{tabular}}

\begin{document}


\if0\blind
{
  \title{\bf Model identification for ARMA time series through convolutional neural networks}
  \author{Wai Hoh Tang \and Adrian R\"ollin}
  \date{National University of Singapore}
} \fi

\if1\blind
{
  \bigskip
  \bigskip
  \bigskip
  \begin{center}
    {\LARGE\bf Model identification for ARMA time series through convolutional neural networks}
\end{center}
  \medskip
} \fi

\maketitle

\begin{abstract}\noindent
We apply convolutional neural networks to the problem of model identification in ARMA time series models, were we train the networks not on actual data, but on simulated time series with known ground truth. Comparing the performance of these networks with traditional likelihood based methods, in particular the Akaike and Bayesian Information Criteria, we are able to show trained networks can significantly outperform likelihood based methods in terms of accuracy and, by orders of magnitude, in terms of speed. We also observe improvements in terms of time series forecasting.
\end{abstract}

\noindent%

\section{Introduction}
\label{sec:intro}

The autoregressive moving average (ARMA) time series model is a classical stochastic model that appears in diverse fields from foreign exchange to biomedical science to rainfall prediction. Using ARMA model for time series analysis typically involves three parts: identification of model orders, estimation of model coefficients, and forecasting. Identification of ARMA orders is crucial, as this has an impact on the subsequent two parts. While the equation describing an ARMA model is simple and easy to interpret, the task of correctly identifying the orders of the autoregressive (AR) and moving average (MA) components to fit a time series is not straightforward. Various methods have been proposed, such as graphical approaches based on autocorrelation function and partial autocorrelation function by \cite{Box1976} and, more commonly, likelihood based methods, such as Akaike information criterion (AIC, see \cite{Akaike1969} and \cite{Durbin2001}), Bayesian information criterion (BIC, see \cite{Akaike1977}, \cite{Rissanen1978} and \cite{Schwarz1978}), Hannan-Quinn information criterion (HQC, see \cite{HannanQuinn1979}), and Minimum Eigenvalue Criterion (MEV, see \cite{Liang1993}).

Artificial intelligence methods such as genetic algorithms (see \cite{Ong2005}, \cite{Palaniappan2006}, \cite{Abo-Hammour2012} and \cite{VanCalster2017}) and artificial neural networks (see \cite{Lee1991}, \cite{Jhee1992}, \cite{LeeJhee1994}, \cite{LeeOh1996},  \cite{Chenoweth2000}, \cite{AlQawasmi2010} and \cite{Jami'in2018}) are other paradigms in model identification. There are common features amongst the methodologies of these artificial neural networks studies, namely limiting the range of ARMA orders, limiting the length of time series and the need for pre-processing of time series. Firstly, the range of ARMA orders in these neural networks papers is typically small; for example, \cite{Chenoweth2000} evaluated orders up to 2 and \cite{LeeOh1996} evaluated orders up to 5. \cite{LeeOh1996} justified that most time series in the real world falls within ARMA(5,5) model. Secondly, the length of time series can vary substantially. \cite{Chenoweth2000} used time series of length 100 and 3,000 while \cite{Jami'in2018} used length 500 and \cite{AlQawasmi2010} used length 1,500. \cite{Chenoweth2000} reasoned that the longer length was chosen to examine an upper limit in accuracy of identification because estimation errors were expected to be minimal for time series of such long length while the shorter length was more representative of real economic data. Thirdly, it is notable that raw time series data are not used directly as inputs in their unprocessed form. Instead, statistical properties or features of time series are used as inputs. \cite{Jami'in2018} used residual error of a recursive least square error algorithm, \cite{AlQawasmi2010} used the special covariance matrix of the MEV criterion as input and \cite{Jhee1992}, \cite{LeeJhee1994} and \cite{Chenoweth2000} used the extended sample autocorrelation function (ESACF) as input. Most papers reckoned that there are promises in using neural networks, and identifications are reasonably accurate but additional work is required for improvement.

In recent years, neural networks have received an increased amount of attention, in particular after \cite{Krizhevsky2012} introduced a convolutional neural network (CNN) architecture, called \emph{AlexNet}, which had won the \emph{ImageNet Large Scale Visual Recognition Challenge} in 2012 and dramatically improved the state-of-the-art of visual recognition and object detection; see the survey by \cite{LeCun2015}. Although the key ideas of the neural network has already been introduced by \cite{LeCun1990}, only advancements in computing power and availability of large data sets have enabled such deep CNN architectures to be built and trained efficiently. Many embellishments of the original architecture have been proposed since; particularly important has been the introduction of so-called \emph{skip connections} by \cite{He2015}, which allowed their architecture \emph{ResNet} to win the ImageNet competition in 2015.

In this paper, we harness the strength of CNNs for the purpose of ARMA time series model selection. At an abstract level, any likelihood-based method can ultimately be seen as a (usually) complicated non-linear function from the data space into some decision space. In our setting, the data space consists of the raw time series, and the decision space the order of AR and MA components predicted by the non-linear function. Our aim will thus be to find a suitable neural network (serving as the non-linear function) that maximise some objective, which, in our case, is the probability of getting the correct ARMA orders. In contrast to earlier artificial neural networks studies, we cover a larger range of ARMA orders. The length of our time series during training is set to 1,000 is in the same order as studies by \cite{Chenoweth2000} and \cite{AlQawasmi2010}. This is also the minimum length of time series that can be processed by our networks because of how the input data shrinks as it funnels through the network due to convolution operation. Once trained, we examine how the networks perform given input time series of different length, namely 1,000, 3,000 and 10,000.

In order to keep the computational load manageable, we have restricted ourselves to a maximal order of both the AR and MA components of 9, amounting to a total of 100 different combinations of ARMA($p,q$) models. This is not a conceptual restriction --- higher orders can easily be achieved by increasing architecture sizes at the cost of also increasing computational time, in particular during training. An important difference in our approach is that, apart from centering and scaling, time series are used directly as inputs to CNNs without prior needs of computing or using any statistical properties of the time series.

The present study stems from a preliminary study\footnote{see \url{https://arxiv.org/abs/1804.04299v1}; the paper elaborates on the selection of CNN architectures and effects of varying different hyper-parameters.}, which has shown promising results. In particular, we have revised our approach in the generation of ARMA time series coefficients by adopting the algorithm of \cite{Beadle1997}, which allows coefficients of ARMA time series to be generated uniformly at random from amongst all admissible coefficients of a given order. We also revised our CNN architecture so that training could be done faster.

The remainder of this article consists of two parts: First, we discuss the mathematical setup and the architecture and training of our CNNs, and second, we compare the neural networks with two classical likelihood-based methods, namely the Akaike and Bayesian Information Criteria. The two likelihood-based methods each come in two flavours: \emph{step-wise} (\acrshort{acr:sw}) and \emph{full} search. We examine the accuracy of ARMA model identification and the resulting impact to time series forecasting.

\section{Methods}\label{sec:meth}

\subsection{Simulation of ARMA(\textit{p}, \textit{q}) time series}\label{SSec:arma}
An ARMA time series model consists of an autoregressive part and a moving average part. We follow the usual convention and denote the AR order by $p$ and the MA order by $q$. A time series $X_t$ following an ARMA($p,q$) model can then be expressed recursively as
\ben{
	X_t = \eps_t + \sum_{i=1}^{p} \phi_i X_{t-i} + \sum_{j=1}^{q} \theta_j \varepsilon_{t-j},
	\label{eq:ARMA}}
where $\phi_1,\dots,\phi_p$ and $\theta_1,\dots,\theta_q$ are real-valued coefficients (typically unknown). $\eps_t$ are independent and identically distributed random values modelling the noise, which in this paper, sampled from standard normal distribution.

Model identification by AIC and BIC selection criteria in \cite{HyndmanKhandakar2008} requires time series to be stationary and invertible, which means all roots of the AR and MA polynomials are larger than 1 (that is, outside the unit circle). Time series used for training of CNN need to be stationary and invertible as well. In fact, the coefficients $\phi_i$ and $\theta_j$ are generated such that the roots are larger than 1.001 to be consistent with the condition applied by \cite{HyndmanKhandakar2008}. In our code, the roots of AR and MA polynomials are determined by constructing companion matrices of the polynomials and solving for corresponding eigenvalues.

The invertibility condition is the counterpart to stationarity for the moving average coefficients, and it implies that the noise can be expressed as a weighted sum of current and past observations; that is, the information in the noise (which is typically not observable) is equivalent to the information of current and past observable data, so that we can write
\ben{
	\eps_t = \sum_{i=0}^{\infty} \pi_i X_{t-i}
}
for some real numbers $\pi_0, \pi_1, \dots$.

While, for example, \cite{Minerva2001}, \cite{Cigizoglu2003} and \cite{Zhang2005} use simulated time series to augment a given dataset in training, we only use simulated data in this paper. Both training and testing data are generated based on~\eqref{eq:ARMA}. We use a standardized length of 1,000 time steps for each time series, excluding the respective burn-in time\footnote{Burn-in period in R \emph{arima.sim} function is computed as $p + q + \ceil{6/\log(m)}$ when $p > 0$, where $m$ is the smallest absolute value of complex roots of the AR polynomial. We modify the last term of burn-in calculation in our code to $\min(50,000, \ceil{10/\log(m)})$. Although this may result in a slight computational overhead, we want to be more stringent. When $p=0$, the burn-in period is set to $q$.}. An effective training of CNN architectures requires a large number of input data and moreover, in order to avoid over-fitting in computer vision tasks, synthetic training data is commonly generated through manipulation of images by cropping, reflecting, etc. Since all our data is simulated and no-time series is seen twice by any CNN, over-fitting cannot occur in our setting --- for the same reason we also do not make use dropout, just like \cite{Ioffe2015} and \cite{He2015}.

\subsection{CNN architectures}\label{SSec:CNNArchitectures}
A basic building module in a ResNet construction is called ResNet block. It comprises of a skip connection parallel to a stack of convolutional weights, batch normalization function and ReLU activation function, which \cite{He2015} calls residual. When data flows through the network, one copy of the data goes through the skip connection path unhindered while an identical copy goes through the residual before merging again in an additive manner at each juncture. We have previously examined different ResNet CNN architectures in \cite{He2016} and established that architecture with \textit{ReLU before addition} construction shows the best training performance. Therefore, this architecture is used in this paper. We train separate neural networks for AR and MA order identification respectively.

Our network consists of one convolutional layer that receives an input data. It applies filters of width 10 and with stride 10, meaning the filter convolves with 10 data points and then moves to the next 10 points. We find that using a stride reduces the amount of memory carried through the network during training. Moreover, training progresses faster. The output of this first layer is 300 feature maps and we use the same number of feature maps for all but the last layer. Filter in subsequent convolutional layers are also of width 10 but move at stride 1. We do not use padding in our setup. From our experiments, padding changes the statistics of the data. When padding is used, validation accuracy of time series of longer length than that used in training, worsens. This should not have been the case since more information is provided. 

\begin{figure}
	\centering
	\includegraphics[width=16cm]{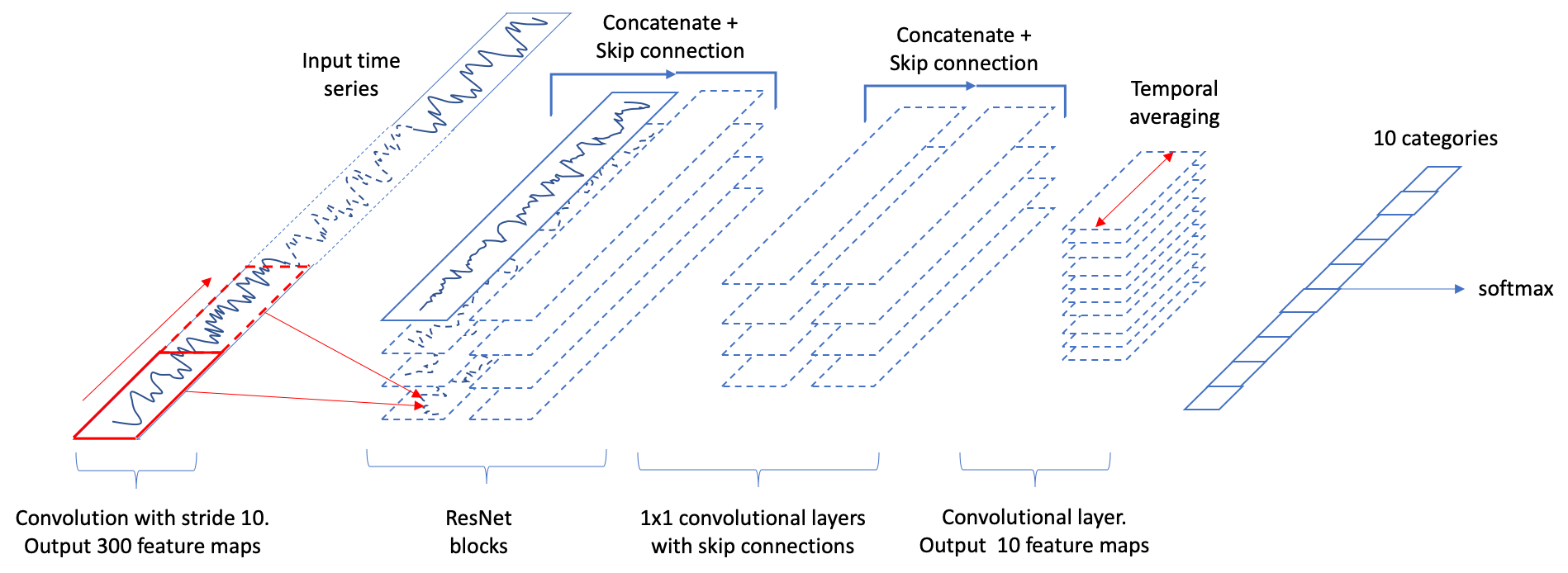}
	\caption{Schematic of our CNN architecture. Data flows from left to right. Filter width is 10 and number of feature maps is 300 for all but last convolutional layer. The first layer convolves the input time series with a stride 10. Subsequent convolutional layers are of stride 1. The data goes through four ResNet blocks and then six $1 \times 1$ convolution with skip connections. The last convolutional layer reduces the 300 feature maps to 10 before a temporal averaging is performed to collapse the values 10 outputs so that softmax can be computed.}
	\label{fig:Architecture}
\end{figure}

Without padding however, the time series shrinks whenever it goes through a convolutional layer. Data flowing through the skip connection have to be truncated at the tail to enable addition to output of residual layers. There are 4 residual blocks of \textit{ReLU before addition} structure used. Shrinkage in time series length limits the number of ResNet block. In order to construct a deeper network, one way to tackle this limitation is to use longer time series for training. However, we add $1 \times 1$ convolution layers instead (see \cite{Lin2013} for details of  $1 \times 1$ convolution). A $1 \times 1$ convolution layer convolves along the `feature dimension', i.e. across feature maps while stride in temporal dimension is 0. It is usually used to downsize (as a bottleneck) or upsize the number of features. In other words, it does not reduce the length of time series passing through it. Leveraging on this property, we arrange them in a `ResNet' like manner, meaning it comes as a pair of $1 \times 1$ convolution layer, batch normalization and ReLU activation function as the residual and connected to a skip connection. There are 6 such blocks. A final convolutional layer convolves along the width dimension again and reduces the number of feature maps to match the number of classes, which is 10 in our study. An averaging along width dimension (i.e. temporal dimension) is done to these ten feature maps before softmax calculation. This averaging step essentially allows our architectures to process longer time series. Figure~\ref{fig:Architecture} shows a schematic of how our network is structured.

The size of a CNN architecture is determined by the total number of tunable parameters in the architecture. Architecture size can be varied by changing the hyper-parameters depth (counted as the number of convolutional layers), filter width and number of features. \cite{He2015} introduced very deep ResNet architectures such as ResNet-50, ResNet-101 and most famously ResNet-152, which won the 2015 ImageNet competition. The prefix digits represent the number of convolutional layers in the respective networks. \cite{He2016} shows that depth of a network improves the expressiveness of the architecture. \cite{Zagorukyo2016} highlighted the large sizes of these ResNet architectures: 25.6, 44.5 and 60.2 million parameters in these three networks respectively. There are no hard-and-fast rules in designing a neural network. \citeauthor{Zagorukyo2016} showed that a wide but shallower network can achieve the same accuracy of a very deep but thin network of comparable size and it is up to 8 times faster to train the former network. Widening of a network is done by increasing the number of feature maps. These ideas motivated us to experiment with a large number of filters and using $1 \times 1$ convolution layers as a mean to increase depth of our networks.

When we started the project, we adapted 2D convolutional neural network for our study. A time series can be perceived as an `image' with 1 channel, height of 1 and width equals to length of the time series. The convolutional weights thus traverse along the width dimension when processing time series. The code could be rewritten for temporal convolution layers, which might be computationally more efficient during training. However, we reckoned our adaptation is equivalent to temporal convolution thus we focus our effort on answering how CNN would fare against information criterion methods.

\subsection{Training of CNN}\label{SSec:CNNtraining}
Training data are generated on the fly during training. One set of training data comprises 100 time series of length 1,000: one time series for each of the 100 possible combinations of $0\leq p\leq 9$ and $0\leq q \leq 9$. For each time series, the coefficients are chosen at random and uniformly based on the algorithm of \cite{Beadle1997}, independently of all other time series, and independently of other batches. The white noise process $\eps_t$ is drawn from a standard normal distribution. Each time series is normalized to zero mean and unit variance. Each batch comprises 12 sets of training data and they are divided to two mini-batches at random. In other words, at each learning step, the network sees 600 time series (per mini-batch). We find using a large batch size helps in the stability of the training. There are 20 batches per epoch.

We use cross-entropy criterion as loss function. While Adam gradient descent is a popular optimizer, our earlier finding shows \emph{Nesterov's Accelerated Gradient} works better for our study. We set the momentum at 0.95 and initial learning rate at 0.1. For one round of training, a network goes through 1,000 epochs with an early stopping condition of 300 continuous epochs without improvement in training error rate. At each epoch, the learning rate decays by 0.995. At the start of each new round, the initial learning rate is reset to 0.99 of the initial learning rate of previous round. The intention is twofold: to reduce learning rate gradually and perturb the system in case the network is stuck at a local minima. We observe that training accuracy improves quickly during initial phase of training before slowing down at about 30\%. At this juncture, we enable the dynamic setting of weights of cross entropy criterion setting, which is discussed in subsection~\ref{SSSec:dynweight}. Our training accuracy progressed to 48.9\% for AR network and 49.3\% for MA network. In an attempt to improve the training  of MA network, we experimented with reducing learning rates and momentum of optimizer and running for more epochs but the improvement was marginal. The eventual MA network registered a training accuracy of 49.6\%.

Training of neural networks are done on GPUs. Forward and backward propagations can be done with matrix multiplications, which are computed efficiently by GPUs. Our GPUs are NVIDIA(R) GeForce(R) GTX 1080 with 8GB of graphic memory.

\subsection{Setting weights of cross entropy criterion dynamically}\label{SSSec:dynweight}
We observe that classification outputs tend to be 0 and 9, which affects the classifications, especially order 1 and 8. Instead of using sampling method or generating different number of time series for different combination of AR and MA order, we leverage on an existing functionality in the neural network framework that we used, which allows us to adjust weights of cross entropy loss. These weights can be used when there is an imbalance in training data but instead, we use them to place different emphasis on corresponding classification orders. According to Torch neural network criterion documentation, for a correct class $c$ and classification probability mass function $x$, the cross entropy loss $L(x,c)$ of a time series at each batch is described as
\ben{
	L(x, c) = w_c \left( -x_c + \log\left( \sum_j  \exp \left( x_j \right) \right) \right),
	\label{eq:CELoss}
}
where $w_c$ is the corresponding weight that can be specified or assigned. By default, these $w_c$'s are all 1's. $x$ is computed automatically by a softmax function at the output layer of the neural network, which allows the outputs of a network to be interpreted as the probability of classifying an input data over a set of classes.

The confusion matrix, which is also known as error matrix, is tracked during the training process. We compute an `analogous' cross entropy loss of each order at each epoch and use that as a proxy to determine the weights, $w_c$'s, of cross entropy loss. The intuition is to set a small weight when the loss of a classification order is low and vice versa when loss is high. In a way, the performance of the network maneuvers itself to emphasize on its `weaker' classifications. Using values in the error matrix, for ground truth class $c$, let $n_j$ be the number of time series that the network classifies as class $j$, where $0\leq j \leq 9$. We compute
\ben{
	y_j = \frac{n_j}{\sum_{j=0}^{9} n_j},
}
and let $\mathcal{L}(y, c)$ be `analogous' cross entropy loss such that
\ben{
	\mathcal{L}(y, c) = - y_c  + \ln \left( \sum_{j=0}^{9} \exp \left( y_j \right) \right),
}In our case, we have 10 classes (i.e. from 0 to 9 shown by index $i$ in the next equation). We compute the mean loss,
\ben{
	\overline{\mathcal{L}} = \frac{1}{10} \sum_{i=0}^{9} \mathcal{L}(y, i).
}
We find that computing `moving average' of 1 period is useful for stability of the training process. Thus at each epoch, the weights of cross entropy loss at time $t$ is
\ben{
	w_{c,t} = 0.5 \times \frac{\mathcal{L}(y, c)}{\overline{\mathcal{L}}} + 0.5 \times w_{c, t-1}.
}

Our approach is akin to cost-sensitive deep learning methods such as \cite{Chung2016} and \cite{Khan2017}, which introduce a cost or penalty element training loss. The approaches in these papers however would require additional coding such as defining a new loss criterion or constructing a new neural network layer, called cost matrix, respectively. Our approach is computationally straightforward.

\section{Validation}
We use the R package \emph{forecast} to generate our test data, to estimate model parameters via AIC and BIC selection criteria, and for forecasting. We generate a test suite of 10,000 time series (i.e.\ 100 batches of 100 time series, one time series for each of the 100 possible combinations of $p$ and $q$ orders from 0 to 9 in each batch) using \emph{arima.sim} function. Similarly, we use \citeauthor{Beadle1997}'s algorithm to sample coefficients of ARMA time series. It is important to emphasize that our test data is generated using readily available code, and is thus \emph{not} the same code as used for training. The white noise is sampled from a standard normal distribution. The length of each time series is 10,010 so that we can examine three cases; using first 1,000, 3,000 and 10,000 points for model identification procedure and subsequent 10 points in each case for forecasting procedure, thus explaining the extra ten points in the time series. By doing this, our aims are firstly to examine the impact of order identifications on forecasting and secondly to observe asymptotic behavior of different methods. The test suite is validated on the AIC and BIC selection criteria as well as on our trained CNN architectures.

The performance of model identification is evaluated in terms of accuracy and computational time. We examine the percentage of correct identifications and \acrfull{acr:mse}, which is the average of squared difference between classified orders and actual orders. A lower \acrshort{acr:mse} means that classified orders are more concentrated around actual values. These two measurements are usually positively correlated. Furthermore, a 95\% confidence intervals of the percentage of correct identifications are estimated by assuming a binomial distribution.

Model identification by AIC and BIC selection criteria is done by \emph{auto.arima} function, in both \textit{step-wise} and \textit{non step-wise}, which we shall term as \textit{full} search. The step-wise algorithm is faster because fewer combinations of $p$ and $q$ are considered. A full evaluation produces better accuracy naturally because all combinations of $p$ and $q$ are considered, but this approach takes much longer. Parameters of the \emph{auto.arima} function are set such that unnecessary computations are avoided\footnote{Setting of some parameters of \emph{auto.arima} function. Order of first-differencing is set to 0, seasonal is set to FALSE, stationary is set to TRUE and allowmean is set to FALSE since the inputs are centred stationary ARMA($p,q$) time series.}, which will maximize processing time without affecting identification accuracy. 

Using the AR and MA orders obtained from the model identification procedure, we proceed with the evaluation of forecasting. The function \emph{arima} is used for fitting and the function \emph{forecast} is used to forecast 1 and 10 periods (i.e. the next 1 and 10 time points in the time  series respectively). We evaluate the performance in terms of \acrfull{acr:mae} and \acrfull{acr:rmse}. Let $N$ be the total number of time series where there is no computer error during forecasting, $T$ be the number periods of forecasting, $X^n_t$ be the actual value of $n$-th time series at time $t$ and $\hat{X}^n_t$ be the forecast. The average performances across these $N$ time series are defined as
\ben{
	\text{\acrshort{acr:mae}} = \frac{1}{N \times T} \sum_{n=1}^{N} \sum_{t=1}^{T} \abs{ \hat{X}^n_t - X^n_t } ,
}

\ben{
	\text{\acrshort{acr:rmse}} = \sqrt{\frac{1}{N \times T} \sum_{n=1}^{N} \sum_{t=1}^{T} \bclr{\hat{X}^n_t - X^n_t }^2},
}

We also conduct a forecasting test where ground truth AR and MA orders are used. We term this test as `Acme' scenario. This test tells us what would the \acrshort{acr:mae} and \acrshort{acr:rmse} of this test suite be, had the model identification method been perfect where all ARMA models are correctly identified. During validation, we noticed a few instances where the forecast function encountered computer errors for some of the time series. This can happen if the original coefficients satisfy the stationarity and invertibility conditions, while the \emph{estimated} coefficients do not, which may happen if the original parameters are very close to non-stationary or non-invertible. Out of 10,000 time series, we were able to process 9,946, 9,967 and 9,968 without any computer errors during forecasting for time series length of 1,000, 3,000 and 10,000 respectively. Furthermore, we performed all our model identification and forecasting on the same machine with a Intel(R) Xeon(R) CPU E5-2620 v4 @ 2.10GHz. In other words, our R code and CNN architectures ran on a CPU during validation. We ran each task on a single-core in order to have a fair comparison of the processing time.

\section{Validation results}
\subsection{Comparison of CNN architectures against information selection criteria}
Table~\ref{tbl:CompareNormalWN} shows results of performance benchmarking. As expected, full evaluations by AIC and BIC selection criteria perform better than step-wise evaluations in terms of identifying AR and MA orders accurately. However, the improvement in accuracy when performing full evaluations comes at a significant increase in computational cost. For example, in the case of time series of length 1,000, the processing time increases from 3.35 hours for BIC step-wise to 48.64 hours for BIC full search and correspondingly from 5.96 hours to 82.09 hours for AIC method. While the improvement from BIC step-wise to full search is significant, it is only marginal in the case of the AIC. The fact that BIC outperforms AIC is consistent with \cite{Hannan1980}. As length of time series increases, processing time goes up considerably. For time series of length 10,000, full evaluations by AIC and BIC methods use 405 and 337 hours respectively. CNN method takes only a fraction of the time to complete the evaluations at 2.35 hours.

For time series of length 1,000, CNN architectures are getting 49.42\% and 49.58\%  of AR and MA orders correct respectively, which are very similar to corresponding training accuracies of 48.9\% and 49.6\%. This indicates that the networks are generalizing well. When tested against increasing length of time series, accuracies increase too, meaning the networks are behaving as expected. It is interesting to note that across the different length of time series, the accuracy of getting both orders correct concurrently is higher than the product of individual AR accuracy and MA accuracy despite the facts that we are using separate networks for each of the order identification and they are trained independently. This therefore suggests that the two networks have `learnt' something in common.

CNN method outperforms AIC and BIC identifying individual AR and MA orders for time series of length 1,000 and 3,000 by a notable margin. Observe however that BIC scales asymptotically better than CNN, especially in terms of getting both orders correct simultaneously. For time series of length 3,000, BIC method scores a 34.27\% accuracy for getting both orders right, edging over 33.80\% by CNN. At length of 10,000, CNN still does better in predicting AR order at 54.03\% accuracy as compared to 50.45\% by BIC full evaluation. In terms of MA order, the accuracy of both methods are comparable and the 95\% confidences intervals overlap.  However, BIC does much better in getting both orders right concurrently at 41.26\% while CNN records a 34.44\%. 

\begin{table}\centering
	\resizebox{\columnwidth}{!}{
		\begin{tabular}{c}
			\multicolumn{1}{c}{Time  series of length 1,000} \\
			\begin{tabular}{@{}rrrrrr@{}}
				\toprule
				& \mline{\tbf{AIC}\\\tbf{\acrshort{acr:sw}}} & \mline{\tbf{AIC}\\\tbf{full}} & \mline{\tbf{BIC}\\\tbf{\acrshort{acr:sw}}} & \mline{\tbf{BIC}\\\tbf{full}} & \mline{\tbf{CNN}} \\
				\midrule
				\tbf{AR(\%)} & 22.41 & 25.69 & 24.44 & 38.53 & 49.42 \\
				\tbf{AR 95\%~C.I.} & [21.60, 23.24] & [24.84, 26.56] & [23.61, 25.29] & [37.58, 39.49] & [48.44, 50.40] \\
				\tbf{AR~\acrshort{acr:mse}} & 10.402 & 8.341 & 9.151 & 5.114 & 1.908 \\[1ex]
				\tbf{MA(\%)} & 23.75 & 29.28 & 23.64 & 40.99 & 49.58 \\
				\tbf{MA 95\%~C.I.} & [22.93, 24.59] & [28.40, 30.18] & [22.82, 24.48] & [40.03, 41.96] & [48.60, 50.56] \\
				\tbf{MA~\acrshort{acr:mse}} & 10.026 & 7.042 & 12.284 & 5.031 & 1.850 \\[1ex]
				\tbf{Both~correct(\%)} & 11.71 & 14.19 & 14.6 & 26.23 & 29.67 \\
				\tbf{Both~correct 95\%~C.I.} & [11.09, 12.36] & [13.52, 14.89] & [13.92, 15.31] & [25.38, 27.10] & [28.78, 30.57] \\[1ex]
				\tbf{Time~taken~(hours)} & 5.96 & 82.09 & 3.35 & 48.64 & 0.30 \\
				\bottomrule
			\end{tabular}
		
			\bigskip
			\\ 
			
			\multicolumn{1}{c}{Time  series of length 3,000} \\
			\begin{tabular}{@{}rrrrrr@{}}
				\toprule
				& \mline{\tbf{AIC}\\\tbf{\acrshort{acr:sw}}} & \mline{\tbf{AIC}\\\tbf{full}} & \mline{\tbf{BIC}\\\tbf{\acrshort{acr:sw}}} & \mline{\tbf{BIC}\\\tbf{full}} & \mline{\tbf{CNN}} \\
				\midrule
				\tbf{AR(\%)} & 24.74 & 27.46 & 28.61 & 45.30 & 53.16 \\
				\tbf{AR 95\%~C.I.} & [23.90, 25.60] & [26.59, 28.34] & [27.73, 29.50] & [44.33, 46.28] & [52.18, 54.14] \\
				\tbf{AR~\acrshort{acr:mse}} & 11.717 & 8.493 & 9.900 & 4.534 & 1.517 \\[1ex]
				\tbf{MA(\%)} & 28.76 & 31.51 & 29.30 & 49.08 & 53.72 \\
				\tbf{MA 95\%~C.I.} & [27.88, 29.66] & [30.61, 32.43] & [28.42, 30.20] & [48.10, 50.06] & [52.74, 54.70] \\
				\tbf{MA~\acrshort{acr:mse}} & 8.745 & 6.984 & 10.565 & 4.251 & 1.525 \\[1ex]
				\tbf{Both~correct(\%)} & 15.24 & 16.86 & 19.75 & 34.27 & 33.80 \\
				\tbf{Both~correct 95\%~C.I.} & [14.55, 15.96] & [16.14, 17.61] & [18.98, 20.54] & [33.35, 35.21] & [32.88, 34.73] \\[1ex]
				\tbf{Time~taken~(hours)} & 17.63 & 179.23 & 11.41 & 119.90 & 0.82 \\				
				\bottomrule
			\end{tabular}
			
			\bigskip
			\\ 
			
			\multicolumn{1}{c}{Time  series of length 10,000} \\
			\begin{tabular}{@{}rrrrrr@{}}
				\toprule
				& \mline{\tbf{AIC}\\\tbf{\acrshort{acr:sw}}} & \mline{\tbf{AIC}\\\tbf{full}} & \mline{\tbf{BIC}\\\tbf{\acrshort{acr:sw}}} & \mline{\tbf{BIC}\\\tbf{full}} & \mline{\tbf{CNN}} \\
				\midrule
				\tbf{AR(\%)} & 26.40 & 28.12 & 33.26 & 50.45 & 54.03 \\
				\tbf{AR 95\%~C.I.} & [25.55, 27.27] & [27.25, 29.01] & [32.34, 34.19] & [49.47, 51.43] & [53.05, 55.01] \\
				\tbf{AR~\acrshort{acr:mse}} & 12.502 & 8.366 & 10.586 & 4.327 & 1.391 \\[1ex]
				\tbf{MA(\%)} & 32.51 & 33.89 & 35.42 & 55.25 & 54.65 \\
				\tbf{MA 95\%~C.I.} & [31.60, 33.43] & [32.97, 34.82] & [34.49, 36.36] & [54.27, 56.22] & [53.67, 55.62] \\
				\tbf{MA~\acrshort{acr:mse}} & 7.820 & 6.668 & 9.223 & 3.899 & 1.434 \\[1ex]
				\tbf{Both~correct(\%)} & 17.30 & 18.37 & 25.67 & 41.26 & 34.44 \\
				\tbf{Both~correct 95\%~C.I.} & [16.57, 18.05] & [17.62, 19.14] & [24.82, 26.54] & [40.30, 42.23] & [33.51, 35.38] \\[1ex]
				\tbf{Time~taken~(hours)} & 63.72 & 405.06 & 44.10 & 337.18 & 2.35 \\
				\bottomrule
			\end{tabular}
		
		\end{tabular}
	}
	\caption{Classifications of ARMA($p,q$) orders for 10,000 time series of length 1,000, 3,000 and 10,000 respectively. CNN generally outperforms information based criterion in identifying the correct AR and MA orders. Both correct are instances where AR and MA orders are correctly identified simultaneously. The \acrshort{acr:mse} values are also smaller, meaning the classifications by CNN are close to actual values. Computational time of CNN is significantly faster.}
	\label{tbl:CompareNormalWN}	
\end{table}

In terms of \acrshort{acr:mse}, CNN records much better results than the other methods and the values improves as the length of time series increases. A smaller \acrshort{acr:mse} means that the identifications by CNN are closer to ground truth on average. Table~\ref{tbl:ARnorm} and \ref{tbl:MAnorm} show detailed breakdown of identification of respective AR and MA orders by BIC full search and CNN. The highest values of each row are located along the diagonals, which shows that both methods identify well across the whole range of $p$ and $q$. Orders identified by CNN are close to the diagonals, illustrating the lower \acrshort{acr:mse} result. These tables show the better performance of CNN, notably for orders higher than 5, for time series of length 1,000 and 3,000 in particular.

\begin{table}\centering
	\resizebox{\columnwidth}{!}{
		\begin{tabular}{c}
			
		\begin{tabular}{c c}	
			\multicolumn{2}{c}{Time  series of length 1,000} \\
			\toprule
			\begin{tabular}{c c}
				& BIC \\
				& Classified AR order  \\
				\rotatebox[origin=c]{90}{Actual AR order} &
				\begin{tabular}{rrrrrrrrrrrr}
					&& \tbf{0} & \tbf{1} & \tbf{2} & \tbf{3} & \tbf{4} & \tbf{5} & \tbf{6} & \tbf{7} & \tbf{8} & \tbf{9} \\
					\midrule
					\tbf{0} && \tbf{\color{red} 61.5} & 16.4 & 6.2 & 3.8 & 1.5 & 1.1 & 1.1 & 2.1 & 2.6 & 3.7 \\
					\tbf{1} && 25.3 &  \tbf{\color{red} 43.1} & 12.2 & 6.2 & 2.6 & 1.9 & 1.7 & 0.9 & 2.5 & 3.6 \\
					\tbf{2} && 10.6 & 13.1 & \tbf{\color{red} 46.8} & 12.8 & 5.2 & 2.8 & 1.7 & 1.4 & 1.8 & 3.8 \\
					\tbf{3} && 6.0 &  8.0 & 22.6 & \tbf{\color{red} 38.1} & 10.3 & 4.5 & 2.5 & 1.5 & 2.4 & 4.1 \\
					\tbf{4} && 3.2 & 2.8 & 12.7 & 17.6 & \tbf{\color{red} 39.1} & 9.2 & 5.1 & 3.3 & 3.5 & 3.5 \\
					\tbf{5} && 1.8 & 4.1 & 8.3 & 11.8 & 19.4 & \tbf{\color{red} 33.0} & 8.4 & 5.1 & 3.7 & 4.4 \\
					\tbf{6} && 1.0 & 2.2 & 4.9 & 5.1 & 14.1 & 18.7 & \tbf{\color{red} 33.2} & 10.3 & 5.6 & 4.9 \\
					\tbf{7} && 0.8 & 1.8 & 2.8 & 5.8 & 10.4 & 11.8 & 21.0 & \tbf{\color{red} 28.5} & 10.5 & 6.6 \\
					\tbf{8} && 0.8 & 0.6 & 3.7 & 4.6 & 7.5 & 7.8 & 16.1 & 17.0 & \tbf{\color{red} 29.8} & 12.1 \\
					\tbf{9} && 0.5 & 1.0 & 1.9 & 4.2 & 7.4 & 7.8 & 11.5 & 14.5 & 19.0 & \tbf{\color{red} 32.2} \\
				\end{tabular}
			\end{tabular}
			
			&
			
			\begin{tabular}{c c}
				& CNN \\
				& Classified AR order  \\
				&
				\begin{tabular}{rrrrrrrrrr}
					\tbf{0} & \tbf{1} & \tbf{2} & \tbf{3} & \tbf{4} & \tbf{5} & \tbf{6} & \tbf{7} & \tbf{8} & \tbf{9} \\
					\midrule
					\tbf{\color{red} 81.2} & 12.0 & 3.9 & 2.2 & 0.3 & 0.2 & 0.1 & 0.1 & 0.0 & 0.0 \\
					27.4 & \tbf{\color{red} 61.2} & 7.2 & 2.5 & 1.2 & 0.5 & 0.0 & 0.0 & 0.0 & 0.0 \\
					10.7 & 19.9 & \tbf{\color{red} 54.7} & 8.6 & 3.8 & 1.0 & 0.8 & 0.4 & 0.1 & 0.0 \\
					5.5 & 8.7 & 25.5 & \tbf{\color{red} 46.8} & 7.8 & 3.7 & 1.2 & 0.5 & 0.2 & 0.1 \\
					1.7 & 3.7 & 9.8 & 23.3 & \tbf{\color{red} 43.7} & 10.9 & 4.4 & 1.3 & 0.9 & 0.3 \\
					1.1 & 1.8 & 4.8 & 12.2 & 22.3 & \tbf{\color{red} 41.4} & 9.8 & 4.5 & 1.1 & 1.0 \\
					0.4 & 0.6 & 2.4 & 4.8 & 11.6 & 20.5 & \tbf{\color{red} 39.0} & 12.4 & 4.9 & 3.4 \\
					0.2 & 0.3 & 1.4 & 2.4 & 6.6 & 13.1 & 21.0 & \tbf{\color{red} 38.8} & 9.3 & 6.9 \\
					0.2 & 0.2 & 1.1 & 1.6 & 3.6 & 6.6 & 11.7 & 19.1 & \tbf{\color{red} 38.0} & 17.9 \\
					0.0 & 0.1 & 0.5 & 0.7 & 1.5 & 3.0 & 6.7 & 13.5 & 24.6 & \tbf{\color{red} 49.4} \\
				\end{tabular}
			\end{tabular}
			\\ \bottomrule				
		\end{tabular}
	
		\bigskip
		\bigskip
		\\
	
		\begin{tabular}{c c}	
			\multicolumn{2}{c}{Time  series of length 3,000} \\
			\toprule
			\begin{tabular}{c c}
				& BIC \\
				& Classified AR order  \\
				\rotatebox[origin=c]{90}{Actual AR order} &
				\begin{tabular}{rrrrrrrrrrrr}
					&& \tbf{0} & \tbf{1} & \tbf{2} & \tbf{3} & \tbf{4} & \tbf{5} & \tbf{6} & \tbf{7} & \tbf{8} & \tbf{9} \\
					\midrule
					\tbf{0} && \tbf{\color{red} 65.3} & 13.6 & 6.5 & 3.4 & 1.1 & 0.9 & 1.1 & 1.3 & 2.1 & 4.7 \\
					\tbf{1} && 19.7 & \tbf{\color{red} 51.5} & 11.0 & 4.5 & 2.1 & 1.6 & 1.1 & 0.9 & 1.9 & 5.7 \\
					\tbf{2} && 7.2 & 12.9 & \tbf{\color{red} 53.0} & 12.0 & 4.0 & 2.5 & 1.7 & 1.2 & 1.7 & 3.8 \\
					\tbf{3} && 3.2 & 5.5 & 18.4 & \tbf{\color{red} 47.7} & 9.9 & 4.8 & 3.0 & 1.5 & 1.7 & 4.3 \\
					\tbf{4} && 1.6 & 2.6 & 8.5 & 17.5 & \tbf{\color{red} 44.4} & 11.2 & 4.7 & 2.2 & 3.6 & 3.7 \\
					\tbf{5} && 2.0 & 2.4 & 6.6 & 8.0 & 21.2 & \tbf{\color{red} 38.8} & 8.3 & 5.5 & 3.2 & 4.0 \\
					\tbf{6} && 0.9 & 2.2 & 3.4 & 4.0 & 11.6 & 16.9 & \tbf{\color{red} 38.7} & 9.7 & 5.8 & 6.8 \\
					\tbf{7} && 0.6 & 1.1 & 2.3 & 4.1 & 8.0 & 10.6 & 21.0 & \tbf{\color{red} 34.8} & 10.4 & 7.1 \\
					\tbf{8} && 0.7 & 0.5 & 2.3 & 4.5 & 5.3 & 7.7 & 12.1 & 15.7 & \tbf{\color{red} 36.4} & 14.8 \\
					\tbf{9} && 0.2 & 0.8 & 1.8 & 3.2 & 4.4 & 6.0 & 8.6 & 13.6 & 19.0 & \tbf{\color{red} 42.4} \\
				\end{tabular}
			\end{tabular}
			
			&
			
			\begin{tabular}{c c}
				& CNN \\
				& Classified AR order  \\
				&
				\begin{tabular}{rrrrrrrrrr}
					\tbf{0} & \tbf{1} & \tbf{2} & \tbf{3} & \tbf{4} & \tbf{5} & \tbf{6} & \tbf{7} & \tbf{8} & \tbf{9} \\
					\midrule
					\tbf{\color{red} 86.0} & 10.4 & 2.8 & 0.7 & 0.0 & 0.0 & 0.1 & 0.0 & 0.0 & 0.0 \\
					25.7 & \tbf{\color{red} 66.2} & 5.9 & 1.7 & 0.3 & 0.1 & 0.1 & 0.0 & 0.0 & 0.0 \\
					9.1 & 21.1 & \tbf{\color{red} 57.8} & 8.7 & 2.3 & 0.7 & 0.2 & 0.0 & 0.0 & 0.1 \\
					3.6 & 8.5 & 25.0 & \tbf{\color{red} 52.7} & 7.2 & 2.2 & 0.6 & 0.1 & 0.1 & 0.0 \\
					1.3 & 2.7 & 9.6 & 22.2 & \tbf{\color{red} 48.7} & 10.3 & 3.8 & 1.0 & 0.2 & 0.2 \\
					0.9 & 1.8 & 4.2 & 10.8 & 22.7 & \tbf{\color{red} 46.0} & 8.8 & 3.0 & 1.6 & 0.2 \\
					0.3 & 0.4 & 2.2 & 3.4 & 10.6 & 22.8 & \tbf{\color{red} 41.1} & 12.6 & 4.7 & 1.9 \\
					0.0 & 0.4 & 0.8 & 1.8 & 5.2 & 12.9 & 22.5 & \tbf{\color{red} 41.3} & 8.5 & 6.6 \\
					0.0 & 0.1 & 0.3 & 1.2 & 2.4 & 6.5 & 12.0 & 20.8 & \tbf{\color{red} 41.3} & 15.4 \\
					0.1 & 0.2 & 0.2 & 0.2 & 1.3 & 2.9 & 5.4 & 13.6 & 25.6 & \tbf{\color{red} 50.5} \\
				\end{tabular}
			\end{tabular}
			\\ \bottomrule				
		\end{tabular}

		\bigskip
		\bigskip
		\\
		
		\begin{tabular}{c c}	
			\multicolumn{2}{c}{Time  series of length 10,000} \\
			\toprule
			\begin{tabular}{c c}
				& BIC \\
				& Classified AR order  \\
				\rotatebox[origin=c]{90}{Actual AR order} &
				\begin{tabular}{rrrrrrrrrrrr}
					&& \tbf{0} & \tbf{1} & \tbf{2} & \tbf{3} & \tbf{4} & \tbf{5} & \tbf{6} & \tbf{7} & \tbf{8} & \tbf{9} \\
					\midrule
					\tbf{0} && \tbf{\color{red} 67.3} & 12.9 & 6.3 & 2.3 & 1.1 & 1.4 & 0.7 & 0.8 & 1.4 & 5.8 \\
					\tbf{1} && 16.5 & \tbf{\color{red} 57.1} & 10.9 & 3.7 & 1.6 & 0.7 & 0.8 & 0.9 & 1.7 & 6.1 \\
					\tbf{2} && 4.9 & 10.3 & \tbf{\color{red} 57.0} & 12.8 & 3.8 & 2.1 & 1.7 & 1.4 & 1.3 & 4.7 \\
					\tbf{3} && 2.2 & 3.9 & 13.8 & \tbf{\color{red} 52.5} & 12.1 & 4.9 & 2.4 & 1.4 & 1.9 & 4.9 \\
					\tbf{4} && 1.2 & 1.7 & 6.3 & 13.7 & \tbf{\color{red} 51.2} & 12.5 & 4.0 & 1.9 & 2.9 & 4.6 \\
					\tbf{5} && 0.9 & 2.5 & 4.1 & 8.2 & 16.7 & \tbf{\color{red} 44.2} & 10.2 & 4.9 & 3.3 & 5.0 \\
					\tbf{6} && 0.4 & 1.8 & 3.1 & 4.0 & 8.1 & 15.3 & \tbf{\color{red} 43.9} & 11.0 & 5.3 & 7.1 \\
					\tbf{7} && 0.5 & 1.0 & 1.7 & 3.5 & 7.2 & 8.3 & 17.4 & \tbf{\color{red} 40.9} & 12.2 & 7.3 \\
					\tbf{8} && 0.7 & 0.7 & 2.9 & 4.3 & 3.2 & 6.4 & 8.8 & 13.5 & \tbf{\color{red} 42.7} & 16.8 \\
					\tbf{9} && 0.5 & 0.7 & 1.8 & 2.7 & 4.9 & 5.1 & 6.1 & 11.5 & 19.0 & \tbf{\color{red} 47.7} \\
				\end{tabular}
			\end{tabular}
			
			&
			
			\begin{tabular}{c c}
				& CNN \\
				& Classified AR order  \\
				&
				\begin{tabular}{rrrrrrrrrr}
					\tbf{0} & \tbf{1} & \tbf{2} & \tbf{3} & \tbf{4} & \tbf{5} & \tbf{6} & \tbf{7} & \tbf{8} & \tbf{9} \\
					\midrule
					\tbf{\color{red} 87.2} & 10.6 & 1.8 & 0.4 & 0.0 & 0.0 & 0.0 & 0.0 & 0.0 & 0.0 \\
					26.7 & \tbf{\color{red} 67.0} & 4.7 & 1.5 & 0.1 & 0.0 & 0.0 & 0.0 & 0.0 & 0.0 \\
					9.0 & 21.4 & \tbf{\color{red} 58.6} & 8.1 & 2.3 & 0.6 & 0.0 & 0.0 & 0.0 & 0.0 \\
					3.7 & 8.9 & 24.5 & \tbf{\color{red} 54.7} & 5.7 & 2.2 & 0.2 & 0.1 & 0.0 & 0.0 \\
					1.2 & 2.8 & 8.8 & 23.0 & \tbf{\color{red} 50.0} & 9.7 & 3.4 & 0.9 & 0.1 & 0.1 \\
					0.6 & 1.7 & 3.5 & 12.5 & 22.0 & \tbf{\color{red} 46.9} & 8.4 & 3.1 & 1.0 & 0.3 \\
					0.3 & 0.5 & 1.6 & 3.2 & 10.8 & 23.9 & \tbf{\color{red} 41.6} & 11.9 & 4.7 & 1.5 \\
					0.1 & 0.2 & 0.8 & 1.9 & 4.6 & 12.7 & 23.2 & \tbf{\color{red} 42.5} & 8.6 & 5.4 \\
					0.0 & 0.2 & 0.1 & 0.9 & 2.0 & 6.6 & 12.2 & 22.4 & \tbf{\color{red} 39.2} & 16.4 \\
					0.0 & 0.0 & 0.2 & 0.0 & 1.0 & 2.9 & 5.2 & 14.6 & 23.5 & \tbf{\color{red} 52.6} \\
				\end{tabular}
			\end{tabular}
			\\ \bottomrule				
		\end{tabular}

		\end{tabular}
	}
	\caption{Classification of AR orders by BIC full evaluation and CNN for time series of length 1,000, 3,000 and 10,000 respectively. Values are in percentages and normalized along the rows. The highest value along each row is highlighted in red.}
	\label{tbl:ARnorm}
\end{table}

\begin{table}\centering
	\resizebox{\columnwidth}{!}{
		\begin{tabular}{c}
			
			\begin{tabular}{c c}	
				\multicolumn{2}{c}{Time  series of length 1,000} \\
				\toprule
				\begin{tabular}{c c}
					& BIC \\
					& Classified MA order  \\
					\rotatebox[origin=c]{90}{Actual MA order} &
					\begin{tabular}{rrrrrrrrrrrr}
						&& \tbf{0} & \tbf{1} & \tbf{2} & \tbf{3} & \tbf{4} & \tbf{5} & \tbf{6} & \tbf{7} & \tbf{8} & \tbf{9} \\
						\midrule
						\tbf{0} && \tbf{\color{red} 82.5} & 7.3 & 3.8 & 1.7 & 1.5 & 0.8 & 0.3 & 0.5 & 0.1 & 1.5 \\
						\tbf{1} && 30.4 & \tbf{\color{red} 52.9} & 7.6 & 3.0 & 1.0 & 1.5 & 0.8 & 0.6 & 0.7 & 1.5 \\
						\tbf{2} && 13.9 & 16.5 & \tbf{\color{red} 51.5} & 7.9 & 2.9 & 2.0 & 1.1 & 0.8 & 0.9 & 2.5 \\
						\tbf{3} && 8.3 & 9.4 & 28.0 & \tbf{\color{red} 39.6} & 6.9 & 2.5 & 1.4 & 0.7 & 1.0 & 2.2 \\
						\tbf{4} && 5.5 & 6.0 & 11.0 & 19.9 & \tbf{\color{red} 41.3} & 7.7 & 3.3 & 2.1 & 1.8 & 1.4 \\
						\tbf{5} && 5.2 & 4.3 & 9.7 & 11.5 & 24.1 & \tbf{\color{red} 29.0} & 7.7 & 3.7 & 1.9 & 2.9 \\
						\tbf{6} && 2.6 & 3.4 & 6.0 & 6.8 & 13.8 & 18.6 & \tbf{\color{red} 33.3} & 7.1 & 4.4 & 4.0 \\
						\tbf{7} && 3.1 & 3.2 & 4.5 & 5.3 & 11.0 & 14.8 & 19.3 & \tbf{\color{red} 25.8} & 7.9 & 5.1 \\
						\tbf{8} && 2.0 & 2.1 & 4.1 & 5.4 & 7.7 & 10.5 & 14.6 & 18.4 & \tbf{\color{red} 26.8} & 8.4 \\
						\tbf{9} && 2.7 & 2.2 & 4.5 & 5.0 & 7.4 & 7.5 & 10.2 & 14.9 & 18.4 & \tbf{\color{red} 27.2} \\
					\end{tabular}
				\end{tabular}
				
				&
				
				\begin{tabular}{c c}
					& CNN \\
					& Classified MA order  \\
					&
					\begin{tabular}{rrrrrrrrrr}
						\tbf{0} & \tbf{1} & \tbf{2} & \tbf{3} & \tbf{4} & \tbf{5} & \tbf{6} & \tbf{7} & \tbf{8} & \tbf{9} \\
						\midrule
						\tbf{\color{red} 83.3} & 10.9 & 3.8 & 1.4 & 0.4 & 0.2 & 0.0 & 0.0 & 0.0 & 0.0 \\
						29.5 & \tbf{\color{red} 59.4} & 7.5 & 2.0 & 0.8 & 0.4 & 0.1 & 0.0 & 0.2 & 0.1 \\
						11.4 & 19.5 & \tbf{\color{red} 53.7} & 9.5 & 3.4 & 1.2 & 0.7 & 0.5 & 0.1 & 0.0 \\
						4.4 & 9.0 & 25.1 & \tbf{\color{red} 47.8} & 8.1 & 3.6 & 0.6 & 0.7 & 0.5 & 0.2 \\
						1.6 & 3.8 & 8.9 & 20.8 & \tbf{\color{red} 45.5} & 12.5 & 3.9 & 1.6 & 0.8 & 0.6 \\
						1.3 & 1.2 & 5.5 & 10.2 & 23.3 & \tbf{\color{red} 41.7} & 9.7 & 4.1 & 1.8 & 1.2 \\
						0.4 & 0.7 & 2.1 & 4.7 & 10.7 & 21.1 & \tbf{\color{red} 39.2} & 12.1 & 6.4 & 2.6 \\
						0.1 & 0.0 & 1.2 & 2.3 & 5.3 & 12.0 & 21.6 & \tbf{\color{red} 37.3} & 12.4 & 7.8 \\
						0.0 & 0.2 & 0.7 & 1.3 & 3.4 & 5.5 & 12.6 & 20.7 & \tbf{\color{red} 38.7} & 16.9 \\
						0.1 & 0.2 & 0.5 & 0.8 & 1.8 & 2.2 & 6.6 & 15.4 & 23.2 & \tbf{\color{red} 49.2} \\
					\end{tabular}
				\end{tabular}
				\\ \bottomrule				
			\end{tabular}
			
			\bigskip
			\bigskip
			\\
			
			\begin{tabular}{c c}	
				\multicolumn{2}{c}{Time  series of length 3,000} \\
				\toprule
				\begin{tabular}{c c}
					& BIC \\
					& Classified MA order  \\
					\rotatebox[origin=c]{90}{Actual MA order} &
					\begin{tabular}{rrrrrrrrrrrr}
						&& \tbf{0} & \tbf{1} & \tbf{2} & \tbf{3} & \tbf{4} & \tbf{5} & \tbf{6} & \tbf{7} & \tbf{8} & \tbf{9} \\
						\midrule
						\tbf{0} && \tbf{\color{red} 84.2} & 5.2 & 1.9 & 1.3 & 1.5 & 0.6 & 0.7 & 0.9 & 0.8 & 2.9 \\
						\tbf{1} && 25.6 & \tbf{\color{red} 60.7} & 6.2 & 1.9 & 1.7 & 0.5 & 0.9 & 0.7 & 0.4 & 1.4 \\
						\tbf{2} && 9.0 & 14.4 & \tbf{\color{red} 60.7} & 7.7 & 2.1 & 1.4 & 1.4 & 0.8 & 0.6 & 1.9 \\
						\tbf{3} && 4.1 & 7.0 & 21.7 & \tbf{\color{red} 49.1} & 9.2 & 2.7 & 2.1 & 0.8 & 0.7 & 2.6 \\
						\tbf{4} && 3.4 & 4.7 & 9.0 & 17.7 & \tbf{\color{red} 48.8} & 6.9 & 4.2 & 1.5 & 1.2 & 2.6 \\
						\tbf{5} && 2.0 & 4.0 & 6.3 & 9.2 & 22.7 & \tbf{\color{red} 39.5} & 7.9 & 3.8 & 1.3 & 3.3 \\
						\tbf{6} && 1.4 & 2.9 & 5.0 & 6.2 & 9.4 & 18.4 & \tbf{\color{red} 39.4} & 8.3 & 5.7 & 3.3 \\
						\tbf{7} && 2.0 & 2.7 & 4.5 & 3.5 & 7.3 & 11.7 & 18.4 & \tbf{\color{red} 35.5} & 8.0 & 6.4 \\
						\tbf{8} && 1.4 & 2.3 & 2.8 & 5.1 & 6.4 & 7.9 & 10.2 & 17.4 & \tbf{\color{red} 35.8} & 10.7 \\
						\tbf{9} && 1.3 & 2.4 & 3.4 & 4.2 & 5.3 & 6.5 & 8.8 & 12.3 & 18.7 & \tbf{\color{red} 37.1} \\
					\end{tabular}
				\end{tabular}
				
				&
				
				\begin{tabular}{c c}
					& CNN \\
					& Classified MA order  \\
					&
					\begin{tabular}{rrrrrrrrrr}
						\tbf{0} & \tbf{1} & \tbf{2} & \tbf{3} & \tbf{4} & \tbf{5} & \tbf{6} & \tbf{7} & \tbf{8} & \tbf{9} \\
						\midrule
						\tbf{\color{red} 87.4} & 9.5 & 2.7 & 0.2 & 0.2 & 0.0 & 0.0 & 0.0 & 0.0 & 0.0 \\
						29.5 & \tbf{\color{red} 63.4} & 5.2 & 1.3 & 0.3 & 0.1 & 0.2 & 0.0 & 0.0 & 0.0 \\
						9.4 & 20.5 & \tbf{\color{red} 59.3} & 7.1 & 2.3 & 0.9 & 0.3 & 0.2 & 0.0 & 0.0 \\
						3.4 & 8.6 & 24.6 & \tbf{\color{red} 54.5} & 5.5 & 2.6 & 0.5 & 0.2 & 0.1 & 0.0 \\
						1.0 & 4.3 & 8.6 & 21.2 & \tbf{\color{red} 49.5} & 10.6 & 2.9 & 1.4 & 0.2 & 0.3 \\
						0.7 & 1.7 & 4.7 & 10.1 & 23.8 & \tbf{\color{red} 45.7} & 8.2 & 3.5 & 1.0 & 0.6 \\
						0.3 & 0.7 & 1.9 & 4.5 & 10.0 & 21.8 & \tbf{\color{red} 40.5} & 12.7 & 5.0 & 2.6 \\
						0.1 & 0.0 & 0.4 & 1.9 & 5.3 & 12.2 & 21.3 & \tbf{\color{red} 42.9} & 9.6 & 6.3 \\
						0.0 & 0.0 & 0.6 & 0.8 & 3.5 & 6.0 & 10.4 & 20.7 & \tbf{\color{red} 43.6} & 14.4 \\
						0.0 & 0.1 & 0.3 & 0.5 & 1.2 & 2.7 & 6.3 & 14.5 & 24.0 & \tbf{\color{red} 50.4} \\
					\end{tabular}
				\end{tabular}
				\\ \bottomrule				
			\end{tabular}
			
			\bigskip
			\bigskip
			\\
			
			\begin{tabular}{c c}	
				\multicolumn{2}{c}{Time  series of length 10,000} \\
				\toprule
				\begin{tabular}{c c}
					& BIC \\
					& Classified MA order  \\
					\rotatebox[origin=c]{90}{Actual MA order} &
					\begin{tabular}{rrrrrrrrrrrr}
						&& \tbf{0} & \tbf{1} & \tbf{2} & \tbf{3} & \tbf{4} & \tbf{5} & \tbf{6} & \tbf{7} & \tbf{8} & \tbf{9} \\
						\midrule
						\tbf{0} && \tbf{\color{red} 86.7} & 4.2 & 1.4 & 1.0 & 1.6 & 0.6 & 0.4 & 0.9 & 0.6 & 2.6 \\
						\tbf{1} && 18.7 & \tbf{\color{red} 68.9} & 5.3 & 1.3 & 1.3 & 0.4 & 0.9 & 0.5 & 0.8 & 1.9 \\
						\tbf{2} && 5.3 & 11.9 & \tbf{\color{red} 66.9} & 6.6 & 2.4 & 1.9 & 0.8 & 0.8 & 0.8 & 2.6 \\
						\tbf{3} && 2.4 & 6.2 & 17.7 & \tbf{\color{red} 55.8} & 8.3 & 3.2 & 1.4 & 1.2 & 0.9 & 2.9 \\
						\tbf{4} && 2.3 & 4.8 & 6.5 & 15.6 & \tbf{\color{red} 54.5} & 6.7 & 3.3 & 1.8 & 1.6 & 2.9 \\
						\tbf{5} && 2.1 & 2.6 & 5.0 & 6.8 & 20.3 & \tbf{\color{red} 46.7} & 7.9 & 3.2 & 1.6 & 3.8 \\
						\tbf{6} && 0.8 & 3.2 & 3.2 & 4.4 & 8.5 & 14.7 & \tbf{\color{red} 47.7} & 7.8 & 5.3 & 4.4 \\
						\tbf{7} && 1.5 & 3.2 & 3.2 & 4.5 & 5.9 & 9.2 & 15.7 & \tbf{\color{red} 39.3} & 10.1 & 7.4 \\
						\tbf{8} && 1.3 & 1.8 & 2.6 & 5.4 & 4.9 & 5.9 & 8.6 & 15.5 & \tbf{\color{red} 42.7} & 11.3 \\
						\tbf{9} && 1.4 & 2.3 & 2.3 & 4.0 & 4.3 & 5.6 & 8.1 & 11.4 & 17.3 & \tbf{\color{red} 43.3} \\
					\end{tabular}
				\end{tabular}
				
				&
				
				\begin{tabular}{c c}
					& CNN \\
					& Classified MA order  \\
					&
					\begin{tabular}{rrrrrrrrrr}
						\tbf{0} & \tbf{1} & \tbf{2} & \tbf{3} & \tbf{4} & \tbf{5} & \tbf{6} & \tbf{7} & \tbf{8} & \tbf{9} \\
						\midrule
						\tbf{\color{red} 89.3} & 8.5 & 1.8 & 0.3 & 0.1 & 0.0 & 0.0 & 0.0 & 0.0 & 0.0 \\
						29.4 & \tbf{\color{red} 63.8} & 4.6 & 1.5 & 0.3 & 0.2 & 0.0 & 0.2 & 0.0 & 0.0 \\
						9.7 & 20.7 & \tbf{\color{red} 59.0} & 7.3 & 2.3 & 0.5 & 0.4 & 0.1 & 0.0 & 0.0 \\
						2.9 & 9.1 & 24.4 & \tbf{\color{red} 55.5} & 4.8 & 2.5 & 0.3 & 0.2 & 0.2 & 0.1 \\
						0.8 & 3.7 & 8.0 & 23.6 & \tbf{\color{red} 50.7} & 9.8 & 2.1 & 0.9 & 0.2 & 0.2 \\
						0.5 & 1.5 & 3.9 & 10.8 & 24.4 & \tbf{\color{red} 46.5} & 8.2 & 2.7 & 1.2 & 0.3 \\
						0.2 & 0.8 & 1.4 & 4.5 & 9.5 & 20.8 & \tbf{\color{red} 43.4} & 12.7 & 4.6 & 2.1 \\
						0.0 & 0.1 & 0.5 & 1.9 & 5.1 & 12.4 & 21.2 & \tbf{\color{red} 43.2} & 10.1 & 5.5 \\
						0.1 & 0.2 & 0.4 & 0.7 & 2.5 & 5.6 & 10.3 & 21.9 & \tbf{\color{red} 43.6} & 14.7 \\
						0.0 & 0.1 & 0.2 & 0.4 & 1.3 & 2.2 & 6.2 & 14.1 & 24.0 & \tbf{\color{red} 51.5} \\
					\end{tabular}
				\end{tabular}
				\\ \bottomrule				
			\end{tabular}
			
		\end{tabular}
	}
	\caption{Classification of MA orders by BIC full evaluation and CNN for time series of length 1,000, 3,000 and 10,000 respectively. Values are in percentages and normalized along the rows. The highest value along each row is highlighted in red.}
	\label{tbl:MAnorm}
\end{table}


\subsection{Time series forecasting based on ARMA models identified}
Table~\ref{tbl:ForecastNormalWN} shows forecast \acrshort{acr:mae} and \acrshort{acr:rmse} when using ARMA models identified by the different methods. In general, forecasts are better when there is a higher accuracy in the identification of ARMA models. This observation supports the importance of identifying the correct ARMA models. Forecast errors increase as the period of forecasting increases, which is expected. Furthermore, forecast results of models identified by AIC and BIC full search are doing better than the step-wise counterparts. On average, ARMA models identified by CNN result in lower \acrshort{acr:mae} and \acrshort{acr:rmse} compared to those identified by AIC and BIC methods. This holds true even for time series of length 10,000 despite not getting both orders correct as good as BIC full evaluations. Interestingly, CNN comes very close to the results of `Acme' scenario, highlighting perhaps the value of getting low \acrshort{acr:mse} of model identifications.

\begin{table}\centering
	\resizebox{\columnwidth}{!}{
	\begin{tabular}{c}
		
		\multicolumn{1}{c}{Time series of length 1,000} \\
		\begin{tabular}{rrrrrrrrrrrrrr}
			\toprule
			& \multicolumn{6}{c}{1 period} & & \multicolumn{6}{c}{10 periods} \\
			& \tbf{Acme} & \mline{\tbf{AIC} \\ \tbf{\acrshort{acr:sw}}} & \mline{\tbf{AIC} \\ \tbf{full}} & \mline{\tbf{BIC} \\ \tbf{\acrshort{acr:sw}}} & \mline{\tbf{BIC} \\ \tbf{full}} & \mline{\tbf{CNN}}
			& & \tbf{Acme} & \mline{\tbf{AIC} \\ \tbf{\acrshort{acr:sw}}} & \mline{\tbf{AIC} \\ \tbf{full}} & \mline{\tbf{BIC} \\ \tbf{\acrshort{acr:sw}}} & \mline{\tbf{BIC} \\ \tbf{full}} & \mline{\tbf{CNN}} \\
			\cmidrule{2-7}  \cmidrule{9-14}
			\tbf{\acrshort{acr:mae}} & 0.8044 & 0.9455 & 0.8449 & 0.9636 & 0.8453 & 0.8052 & & 1.7905 & 2.0395 & 1.9110 & 2.0635 & 1.9081 & 1.7959 \\
			\tbf{\acrshort{acr:rmse}} & 1.0101 & 1.8427 & 1.2002 & 1.8563 & 1.2001 & 1.0123 & & 3.1187 & 4.0823 & 3.6017 & 4.1059 & 3.5945 & 3.1708 \\
			\bottomrule
		\end{tabular}
		
		\bigskip
		\\
		
		\multicolumn{1}{c}{Time series of length 3,000} \\
		\begin{tabular}{rrrrrrrrrrrrrr}
			\toprule
			& \multicolumn{6}{c}{1 period} & & \multicolumn{6}{c}{10 periods} \\
			& \tbf{Acme} & \mline{\tbf{AIC} \\ \tbf{\acrshort{acr:sw}}} & \mline{\tbf{AIC} \\ \tbf{full}} & \mline{\tbf{BIC} \\ \tbf{\acrshort{acr:sw}}} & \mline{\tbf{BIC} \\ \tbf{full}} & \mline{\tbf{CNN}}
			& & \tbf{Acme} & \mline{\tbf{AIC} \\ \tbf{\acrshort{acr:sw}}} & \mline{\tbf{AIC} \\ \tbf{full}} & \mline{\tbf{BIC} \\ \tbf{\acrshort{acr:sw}}} & \mline{\tbf{BIC} \\ \tbf{full}} & \mline{\tbf{CNN}} \\
			\cmidrule{2-7}  \cmidrule{9-14}		
			\tbf{\acrshort{acr:mae}} & 0.8051 & 0.9434 & 0.8364 & 0.9566 & 0.8363 & 0.8059 & & 1.8033 & 2.0393 & 1.9182 & 2.0584 & 1.9180 & 1.8086 \\
			\tbf{\acrshort{acr:rmse}} & 1.0089 & 1.8332 & 1.0789 & 1.8463 & 1.0794 & 1.0108 & & 3.1536 & 4.1655 & 3.5462 & 4.2002 & 3.5437 & 3.1616 \\
			\bottomrule
		\end{tabular}
	
		\bigskip
		\\
		
		\multicolumn{1}{c}{Time series of length 10,000} \\
		\begin{tabular}{rrrrrrrrrrrrrr}
			\toprule
			& \multicolumn{6}{c}{1 period} & & \multicolumn{6}{c}{10 periods} \\
			& \tbf{Acme} & \mline{\tbf{AIC} \\ \tbf{\acrshort{acr:sw}}} & \mline{\tbf{AIC} \\ \tbf{full}} & \mline{\tbf{BIC} \\ \tbf{\acrshort{acr:sw}}} & \mline{\tbf{BIC} \\ \tbf{full}} & \mline{\tbf{CNN}}
			& & \tbf{Acme} & \mline{\tbf{AIC} \\ \tbf{\acrshort{acr:sw}}} & \mline{\tbf{AIC} \\ \tbf{full}} & \mline{\tbf{BIC} \\ \tbf{\acrshort{acr:sw}}} & \mline{\tbf{BIC} \\ \tbf{full}} & \mline{\tbf{CNN}} \\
			\cmidrule{2-7}  \cmidrule{9-14}			
			\tbf{\acrshort{acr:mae}} & 0.7891 & 0.9710 & 0.8234 & 0.9773 & 0.8239 & 0.7900 & & 1.7872 & 2.0846 & 1.9125 & 2.0931 & 1.9141 & 1.7923 \\
			\tbf{\acrshort{acr:rmse}} & 0.9875 & 4.4579 & 1.1198 & 4.4607 & 1.1200 & 0.9886 & & 3.0160 & 5.6511 & 3.5758 & 5.6577 & 3.5811 & 3.0481 \\
			\bottomrule
		\end{tabular}	
	
	\end{tabular}

	}
	\caption{Forecasting for 1 and 10 periods. Number of time series without computer error were 9,946, 9,967 and 9,968 for time series of length 1,000, 3,000 and 10,000 respectively. `Acme' scenario is forecasting done with ground truth ARMA orders, i.e. all ARMA models are correctly identified. Forecasting using ARMA models identified by CNN achieves the lower error rates on average compared to models identified by AIC and BIC methods in all cases.}
	\label{tbl:ForecastNormalWN}	
\end{table}

\clearpage 
\section*{Acknowledgements}

The authors thank Ying Chen and Alexandre Thiery for helpful discussions. This project was partially supported by MOE Tier 1 Research Grant R-155-000-213-114.


%
%
%

{}

\end{document}